# Achieving highly strengthened Al-Cu-Mg alloy by grain refinement and grain boundary segregation

Takahiro Masuda[1][※], Xavier Sauvage[2], Shoichi Hirosawa[1] and Zenji Horita[3,4,5]

[1] Department of Mechanical Engineering and Materials Science, Yokohama National University, Yokohama 240-8501, Japan

[2] Normandie Univ, UNIROUEN, INSA Rouen, CNRS, Groupe de Physique des Matériaux, 76000 Rouen, France

[3] Graduate School of Engineering, Kyushu Institute of Technology, Kitakyushu 804-8550, Japan

[4] Magnesium Research Center, Kumamoto University, Kumamoto, 860-8555, Japan

[5] Synchrotron Light Application Center, Saga University, Saga, 840-8502 Japan

※Corresponding author: masuda-takahiro-fk@ynu.ac.jp

**Abstract**

An age-hardenable Al-Cu-Mg alloy (A2024) was processed by high-pressure torsion (HPT) for producing an ultrafine-grained structure. The alloy was further aged for extra strengthening. The tensile strength then reached a value as high as ~1 GPa. The microstructures were analyzed by transmission electron microscopy and atom probe tomography. The mechanism for the high strength was clarified in terms of solid-solution hardening, cluster hardening, work hardening, dispersion hardening and grain boundary hardening. It is shown that the segregation of solute atoms at grain boundaries including subgrain boundaries plays a significant role for the enhancement of the tensile strength.

**Keywords:** aluminum alloys, severe plastic deformation, grain refinement, high strength, grain boundary segregation

# 1. Introduction

Since the discovery of age hardening in an Al-Cu-Mg alloy by Wilm in 1906 [1], age-hardenable aluminum alloys, such as 2xxx series (Al-Cu-Mg system), 6xxx series (Al-Mg-Si system) and 7xxx series (Al-Zn-Mg-Cu system), have been extensively studied for wide ranges of compositions. These alloys are used as high-specific-strength structural materials for aircraft and automotive industries [2,3]. These commercial aluminum alloys usually rely on two strengthening contributions, namely solute atoms and precipitates, but do not take advantage of grain boundary (GB) strengthening since they exhibit coarse-grained structures.

It is well known that the strength of metallic materials also increases with the reduction in grain size through the Hall-Petch relation [4,5]. The grain size is easily reduced to the submicron level when processes of severe plastic deformation (SPD) are utilized [6]. Main SPD processes are ECAP (equal-channel angular pressing) [7], HPT (high-pressure torsion) [8], ARB (accumulative roll bonding) [9], MDF (multi-directional forging) [10] and HPS (high-pressure sliding) [11]. It has been reported that the grain size of pure aluminum (99.99%) can be reduced to ~1 μm by ECAP, HPT, ARB and HPS [11-14]. It typically gives rise to a significant increase in the yield stress following the Hall-Petch law [4,5]. This strategy can also be applied to commercial Al alloys [15-25]. Liddicoat *et al*. reported that the tensile strength increased to > 1 GPa in an ultrafine-grained (UFG) A7075 alloy processed by HPT [15]. It was considered that this high strengthening is due to non- equilibrium structures such as the formation of solute clusters and the segregation of solute atoms along GBs. Significant strengthening was also reported for UFG A2024 alloys processed by SPD [18,22]. Interestingly, the precipitation sequence usually takes a different type, with the easy nucleation of stable phase without the classical sequence (through formation of Guinier-Preston-Bagarysatsky (GPB) zones [26]). Besides, strain-induced segregations of solute atoms along GBs have been reported in the Al-Cu system [27, 28] and other alloys [29]. In addition, three dimensional atom probe tomography (APT) by Sha *et al*. [30] showed that such segregation of Cu and Mg also occurred during aging in an A2024 alloy. They could efficiently stabilize UFG structures but also contribute to strengthening [29].

The aim of this research is threefold: first, to achieve ultra-high strengthening to a level as high as ~1 GPa in an age-hardenable A2024 alloy. HPT process is applied for adequate grain refinement, and subsequent aging is undertaken for precipitation and segregation of solute atoms. Second, to evaluate accurately microstructural features down to the nanoscale with a special emphasis on solute atom distribution using scanning transmission electron microscopy (STEM) and APT. Third, to clarify strengthening mechanisms thanks to a quantitative evaluation of strengthening contributions. We will then show that the ultrahigh strength achieved in an A2024 alloy is not only controlled by grain refinement but also by segregation of solute atoms along GBs including subgrain boundaries.

## 2. Experimental procedures

Disk samples with 10mm diameter were prepared from a 1mm thickness commercially available A2024 alloy sheet with the composition of Al-4.5Cu-1.4Mg-0.56Mn-0.07Fe-0.02Si-0.01Ti (in wt.%). The disks were subjected to solution treatment (ST) at 793 K for 1 h followed by water quenching. The average grain size for the ST sample was determined to be 62 μm.

Samples were processed by HPT at room temperature (R.T.) for 0.75, 1, 5 and 10 revolutions with a rotation speed of 1 rpm under a pressure of 6 GPa. The imposed strain introduced by HPT processing can be represented using equivalent strain ($\varepsilon_{eq}$) through the following equation:

$$\varepsilon_{eq} = 2\pi rN/\sqrt{3}t.\qquad(1)$$

where, $r$ is the distance from the disk center, $N$ is the number of revolutions and $t$ is the thickness of the disk after HPT processing. Thus, the equivalent strain introduced in each sample by HPT processing was determined to be 6, 8, 26, 44 and 88 for 0.75, 1, 3, 5 and 10 revolutions, respectively, at the gauge area (corresponding to $r$ = 2 mm) where the thickness was taken after HPT processing as 0.8~0.85 mm. Following HPT processing, some of the samples were subsequently aged at 423 K in an oil bath for 35 min and 3 h.

Samples processed by HPT and post-HPT aging were ground to the mid-plane of the thickness by abrasive papers and polished to a mirror-like surface with an alumina suspension. Hardness measurements were made along 8 different radial directions with an interval of 0.5 mm on the same line as shown in Fig. 1. For the hardness measurements, a load of 0.3 kgf was applied for 15 s using a Mitutoyo HM-102 tester.

The dimensions of the tensile specimens are also depicted in Fig. 1 (gauge part: 1.5mm long, 0.6mm wide, 0.6mm thick) and the specimens were extracted at 2 mm away from the disk center. Tensile tests were performed at R.T. with an initial strain rate of $3.0 \times 10^{-3}$ s$^{-1}$. Tensile tests were repeated at least four time for each condition.

Microstructural observations after HPT processing and post-HPT aging were conducted on the plane perpendicular to the rotation axis using transmission electron microscopy (TEM), scanning transmission electron microscopy (STEM) and energy Dispersive X-ray Spectroscopy (EDS) analyses. For preparation of electron-transparent thin specimens for TEM, disks with 3mm diameter were extracted from HPT-processed samples as shown in Fig. 1 and they were ground on both sides with abrasive papers to a thickness of 0.15 mm. These disks were further thinned using a twin-jet electrochemical polisher in a solution of 20% $HNO_3$-80% $CH_3OH$ (in vol.%) at a temperature of 253 K under 10 V. Further thinning was carried out by low-energy ion milling conducted with a GATAN® Precision Ion Polishing System (accelerating voltage: 3 kV, incidence angle of Ar ions: ±3°, duration: 3 min). Observations were performed using a Hitachi-8100 and a JEOL-ARM200F microscopes operated at 200 kV. The average grain size was determined from more than 100 grains imaged in dark-field mode.

Second phase identification and dislocation density measurements were carried out by X-ray diffraction (XRD) analyses using CuKα radiation with a scanning speed of 0.2°/min and a scanning step of 0.01°.

APT analyses were carried out with a CAMECA LEAP 4000HR instrument. Samples were prepared by a classical two step electro-polishing method starting from small rods (0.3×0.3 $mm^2$ in cross-section) cut at a distance of 3 ± 1 mm from the disk center. Sharpening of tips was done with a solution of 2% perchloric acid in ethylene glycol monobutyl ether at R.T. under an applied voltage of 20 V. Samples were field-evaporated at 30 K with electric pulsing (pulse repetition rate 200kHz, pulse fraction 20%). Data processing was performed using "GPM 3D data software". It should be noted that the analyses for TEM, STEM, XRD and APT were conducted in the areas well away from the disk center after $N$=10 and/or subsequent aging, where the microstructure including the grain size is considered to be the same and uniform according to an earlier report [13].

## 3. Experimental results
### 3.1 Tensile tests and hardness measurements

Fig. 2 shows (a) representative stress-strain curves and (b) the tensile strength and elongation to failure plotted as a function of the number of revolutions for the samples processed by HPT through $N$=0.75 to $N$=10. Fig. 2(a) also includes, for comparison, the results after ST and standard T6 aging (at 463 K for 12 h) for the A2024 alloy. Fig. 2 shows that processing through $N$=0.75 led to an increase in the strength to 750 MPa which is already 50% higher than that of the T6 sample. The tensile strength was further increased with increasing $N$ and it reached 910 MPa with the total elongation of 5% after $N$=10. It should be noted that the higher tensile strength was achieved as compared to an earlier study (890 MPa) [22]. This may be attributed to the larger imposed strain ($\varepsilon_{eq}$ =88 against 13 in [22]).

Fig. 3 shows (a) Vickers microhardness plotted against equivalent strain and (b) stress-strain curves after tensile testing at R.T., where the samples were processed by HPT through $N$=10 and subsequently aged at 423 K for 35 min. It should be noted that the aging period for 35 min corresponds to peak aging as reported before [22]. Fig. 3 also includes the strength level for the ST and T6 samples. Fig. 3(a) clearly shows that the hardness was further increased up to 285 HV during aging, appreciably above the hardness of 260 HV reached after HPT processing. The results after tensile testing in Fig. 3(b) confirm that the mechanical strength has increased during aging, with a tensile strength up to 1 GPa. Thus, the HPT-processed sample combined with subsequent aging exhibits the tensile strength twice as high as the sample after T6 treatment.

**3.2 Grain size and dislocation density**

Fig. 4 shows TEM bright-field images (left), dark-field images (right) and the corresponding SAED patterns (inset) recorded with an aperture size of 1.3 μm for (a) the as-HPT-processed ($N$=10) sample and (b) the subsequently peak-aged sample. Each dark-field image was taken using the diffracted beam indicated by the arrow in the corresponding SAED pattern. The TEM observations revealed that the grain size was refined to 130 nm. The SAED pattern exhibits Debye-Scherrer rings indicating that the grains are mostly surrounded by high-angle grain boundaries (HAGBs) (Fig. 4(a)). Thus, a significant grain refinement was achieved from the initial grain size of 62 μm. A slight grain growth occurred during the aging but the UFG structure was kept with a mean grain size of 160 nm (Fig. 4(b)). The UFG structures observed in this study are consistent with the results reported by Mohamed *et al*. [18], Alhamidi *et al*. [31], Chen *et al*. [32, 33] and Dobatkin *et al*. [34] with the grain sizes of 240 nm, 240 nm, 160 nm and 150 nm, respectively.

Fig. 5 shows XRD profiles of the samples after $N$=10 and subsequent aging at 423 K for 35 min (peak-aged) and 3 h (over-aged). The XRD profile for the ST sample is also included for comparison. These profiles, except for the over-aged sample, do not show any peak that could be attributed to Al$_2$Cu ($\theta$) or Al2CuMg ($S$) phases. A small volume fraction of intermetallic particles, identified as Al$_{20}$Cu$_2$Mn$_3$ dispersoids (by STEM-EDS analysis, data not shown here), was detected. An extra peak corresponding to the stable $S$ phase also appeared in the over-aged condition (*e.g.*, in the vicinity of 35º). This is consistent with earlier works carried out on a similar material processed by SPD where the $S$ phase was also observed in over-aged conditions [18, 22].

Dislocation densities were estimated from peak broadening using the Williamson-Hall method [35, 36]. The calculation was carried out using five different full widths at half maximum (FWHMs) obtained from (111), (200), (220), (311) and (222) planes after normalization by calibrating the instrument [37]. The dislocation density measured in the as-HPT-processed sample was $5.5 \times 10^{15}$ m$^{-2}$ and it decreased to $1.0 \times 10^{15}$ m$^{-2}$ after peak-aging.

### 3.3 Microstructures

Fig. 6 shows STEM images obtained in (a, b) the as-HPT-processed sample and (c, d) the subsequently peak-aged sample where BF images are in (a) and (c) and the corresponding HAADF images in (b) and (d). In the BF images, crystalline defects create some contrast variations within a typical length scale much smaller than the grain size. From the corresponding HAADF images, some changes in local contrast are visible as arrowed, indicating that copper (the element with the largest atomic number in the A2024 alloy) has segregated along these boundaries. Thus, these boundaries are mainly attributed to the presence of low-angle grain boundaries (LAGBs). The segregations are more pronounced in the sample after peak-aging than in the as-HPT-processed state. As depicted in the HAADF line profile (Fig. 6(e)), the typical thickness of such a Cu rich layer is 2 nm or less. EDS confirmed the enrichment of Cu and Mg solutes at GBs (Fig. 6(f)). The mean distance between such boundaries ($\lambda$) with enhanced local Cu and Mg concentration was estimated from the HAADF images. It should be noted, however, that the measurement of $\lambda$ was difficult for the as-HPT-processed sample because the fraction of the segregated boundaries was low ($\leqq 0.1$) and thus contrast was poor (Fig. 6(b)). After aging, the solute enrichment is appreciable at almost all boundaries, so that $\lambda$ could be estimated as 40±10 nm.

Fig. 7 shows the three-dimensional atom distribution maps of (a) Al, Cu and Mg, (b) Cu and (c) Mg obtained by APT for the as-HPT-processed sample. A GB is visible at the center, where the segregation of Cu and Mg is evident. The segregation is well demonstrated from the compositional profiles across the GB shown in Fig. 7(d). The concentrations of Cu and Mg are significantly increased at the boundary which is fully consistent with STEM-EDS data (Fig. 6(f)).

The distributions of Cu and Mg at the nanoscale were also investigated by APT after subsequent peak-aging. Intensive segregations along two-dimensional defects are clearly visible in Fig. 8(a) and (b) with the distributions of Cu and Mg respectively. Interestingly, few precipitates could be observed within the interior of the grains surrounded by the segregated boundaries, despite the analyzed volume which was large enough to cover more than several grains. This was also confirmed with repeated APT analysis. The APT analysis was further repeated with two more samples in the same peak-aged state. Essentially, the same conclusion is reached as documented in Fig. 9. The intensive segregations of Cu and Mg are confirmed in Fig. 9 (a) and (b). In Fig. 9(c), the data was filtered with a 4at.% threshold for a better visualization of solute rich regions. This filtering analysis reveals that the typical length scale of these segregations stands in a range of 20 to 50 nm which is considerably smaller than the grain size of ~160 nm determined for grains with HAGBs in Fig. 4(b) and in agreement with STEM-HAADF observations (Fig. 6(d)). This finding confirms that segregations occurred not only at HAGBs but also at LAGBs. To check any presence of precipitates within the grains, the data were filtered with a higher composition threshold (15at.%Cu) (Fig. 9(d)). Only two Cu-rich nanoscaled precipitates could be detected as circled in Fig. 9(d). One appears elongated (about 5 nm in diameter and 20 nm long), the other is spherical (about 5 nm in diameter), and both contain about 22±7 at.%Cu. Thus, they correspond to the $Al_2Cu$ ($\theta'$) precipitates even if the Cu content is slightly lower than expected for this phase, as reported in other studies [38, 39].

Then, to determine if some nanoscaled precipitates have nucleated during aging at 423 K, frequency distribution histograms of Cu and Mg atoms located between segregated boundaries were calculated and compared to histograms where atom positions were randomized (Fig. 10(a,b)). These histograms clearly show that Cu and Mg atoms in the solid solution are randomly distributed both in the as-HPT-processed and in the subsequently peak-aged states. Thus, only atomic clusters [32, 33] should exist in these two states.

Concentration profiles (such as in Fig. 10(c)) were computed to quantify the segregated Cu and Mg levels. They are relatively similar to that measured in the as-HPT-processed state (Fig. 7(d)). The mean solute excesses at the segregated boundaries correspond to about one eighth of Cu monolayer and one eighth of Mg monolayer. The difference of composition between the whole volume and the concentration measured in regions along boundaries with segregations (Table 1) shows that about 0.5at.% of solutes has segregated to the boundaries. As GBs being shared between two grains, the fraction of atomic sites at boundaries ($f_{s\_GB}$) can be written as:

$$f_{s\_GB} = S\, V_{at}^{\frac{1}{3}}/2V \;, \qquad (2)$$

where $S$ is the grain surface area, $V$ is the grain volume and $V_{at}$ is the mean atomic volume ($V_{at} \sim 0.016$ nm$^3$ in fcc Al). For spherical grains with a diameter of 150 nm, a monolayer of solute is achieved with 0.5at.% of solute. This is much larger than the length scale measured by APT (Fig. 9) and seen in the HAADF images (Fig. 6). However, in the corresponding images, it should be noted that boundaries exhibit an elongated shape with average dimensions close to 100x100x20 nm$^3$. Such dimensions give a decrease in solute to ~0.5at.% to cover boundaries with one quarter of Cu and Mg monolayer (from eq.(2)).

## 4. Discussion
### 4.1 Microstructural features

Fig. 11 illustrates the summary of microstructural features based on the results by XRD, TEM, STEM and APT. In the as-HPT-processed state ($N$=10) (Fig. 11(a)), the UFG structure with the grain size of 130 nm containing a high density of LAGBs is well developed, and XRD shows much distortion due to a high density of dislocations. Intriguingly, the solute segregation was observed along a few boundaries but the fraction of this segregation is as small as ~10%. The segregation occurred throughout all types of GBs including high- and low-angles of misorientations after subsequent peak-aging (Fig. 11(b)), leading to the extra reduction in the mean distance between segregated boundaries to 20-50 nm as is evident from the APT analysis, while the amount of precipitates is negligibly small as observed by APT. It should be also noted that the ultrafine-grained structure is well retained after aging ($d$ = 160 nm) owing to the strong solute segregation onto GBs, although a significant reduction in dislocation density occurred due to recovery.

**4.2 Strengthening mechanism**

In this study, the strength of the A2024 alloy reached 1 GPa and it is important to clarify the mechanism for this notable strengthening. The factors affecting this strengthening can be due to the interaction of dislocations with solute atoms and clusters, the mutual interaction of dislocations, the pinning of dislocations by dispersed particles and the blocking of dislocations by GBs, each of which corresponds to, so called, solid-solution hardening ($\Delta\sigma_{ss}$), cluster hardening ($\Delta\sigma_{cluster}$), work hardening ($\Delta\sigma_{dis}$), dispersion hardening ($\Delta\sigma_{pre}$) and grain boundary hardening ($\Delta\sigma_{gb}$), in addition to overcoming the Piers potential ($\sigma_o$). The total strength may then be described by the summation of all the strengthening factors [40].

$$\sigma_{0.2} = \sigma_0 + \Delta\sigma_{ss} + \Delta\sigma_{dis} + \Delta\sigma_{pre} + \Delta\sigma_{cluster} + \Delta\sigma_{gb}. \tag{3}$$

Here, $\sigma_{0.2}$ is the yield strength and $\sigma_o$ is the friction stress equivalent to the Piers stress.

The values of the yield stress were measured as $\sigma_{0.2}$ =770 MPa and 955 MPa for the as-HPT-processed sample and the subsequently peak-aged sample, respectively, from the stress-strain curves in Fig. 3(b). The strength increased during aging despite grain growth and the decrease in dislocation density as illustrated in Fig. 11. Numerical estimation of each strengthening factor is derived as follows.

### 4.2.1 Evaluation of strengthening factors

*Friction stress*: The value of $\sigma_o$ is taken from the yield stress reported in pure Al (99.999 %) so that $\sigma_o$ = ~5 MPa [41].

*Solid-solution hardening*: The value of $\Delta\sigma_{ss}$ is estimated from the Fleischer [42] and the Labusch [43] equations so that $\Delta\sigma_{ss}$ =~50 MPa. Here, this estimation is due to the average of the both values predicted from the Fleischer and the Labusch equations, which are 37 MPa and 59 MPa, respectively.

*Work hardening*: For the estimation of $\Delta\sigma_{dis}$, the following Bailey-Hirsch equation [44] is used:

$$\Delta\tau_{dis} = \alpha G b \sqrt{\rho} \ . \tag{4}$$

The dislocation density is input from the values measured from the XRD profile shown in Fig. 5, and thus it follows that $\Delta\sigma_{dis}$ =406±11 MPa for the as-HPT-processed state, but $\Delta\sigma_{dis}$ decreases to 174±8 MPa after the subsequent peak-aging. This calculation has used α=0.24 for Al [45] and the Taylor factor (*M*=3.06 [46-48]).

*Precipitation hardening*: The contribution of $\Delta\sigma_{pre}$ is considered to be zero as is evident from the APT analysis that the presence of the $\theta$' phase is negligible.

*Cluster hardening*: For the value of $\Delta\sigma_{cluster}$, it was estimated from tensile testing of the sample after ST and subsequent aging at 423 K for 35 min which corresponds to the peak-aged condition for the HPT-processed sample. It was confirmed that the values of yield stress for each sample are 175 MPa and 280 MPa as $\sigma_{y(ST)}$ and $\sigma_{y(aged)}$, respectively, but these include the contributions of $\sigma_o$, $\Delta\sigma_{ss}$ and $\Delta\sigma_{gb}$. Thus, $\Delta\sigma_{cluster}=\sigma_y-(\sigma_o+\Delta\sigma_{ss}+\Delta\sigma_{gb})$= ~110 MPa (for $\sigma_{y(ST)}$) and ~215 MPa (for $\sigma_{y(aged)}$). (Note that $\sigma_{gb}$ was estimated as 10 MPa from the Hall-Petch equation.) Sha *et al.* [30] reported that the hardness of a solution-treated

A2024 alloy increased by aging at 443 K for 30 min although no precipitation was detected. Instead, they concluded that the hardness increased due to the formation of clusters. It should be noted that the cluster formation in the HPT-processed sample may not be the same way as the ST sample. Nevertheless, the APT analysis clearly indicated that the clusters are present as shown in Fig. 9(d).

*Grain boundary hardening*: The value of $\Delta\sigma_{gb}$ is estimated using the following Hall-Petch equation [4,5]:

$$\Delta\sigma_{gb} = \sigma - \sigma_0 = \frac{k_y}{\sqrt{d}}, \quad (5)$$

where $k_y$ is a constant depending on the alloy type and composition. In this study, $k_y$ =0.08 MPa m$^{1/2}$ was used from the report by Shanmugasundaram in an Al-4%Cu alloy where the effect of dislocations and precipitations were minimized [49]. As observed in Fig. 4(a) and (b), the grain sizes surrounded by high-angle boundaries were of 130 nm and 160 nm for the as-HPT-processed and the subsequently peak-aged samples, respectively. These grain sizes were then adopted for $d$ in the above Hall-Petch equation, and thus it follows that $\Delta\sigma_{gb(HAGB)}$ = 222±40 MPa and 200±50 MPa, respectively.

The comparison between the measured value and the estimated value ($\Delta\sigma_y$ (with $\Delta\sigma_{gb(HAGB)}$)) in Table 2 shows that the estimated strength is almost comparable to the measured one for the as-HPT-processed sample, but the difference is significant for the sample subjected to aging after HPT processing (Table 2). In order to explain this difference, we consider the additional contribution to the strengthening: the segregation of solute elements onto GBs including not only high-angle misorientations but also low-angle misorientations as observed by APT (Fig. 8, 9).

**4.2.2 Segregation on low-angle grain boundaries**

Based on the analytical results by APT that Cu and Mg were segregated on GBs by 5 at% on average, it can be estimated if the presence of such solute atoms is effective to stabilize the GBs in terms of the local strain developed by dislocations. Here, we considered that the GB is low-angle tilt boundary with an inclination of 10 degrees, which is typical of LAGBs. Provided that the angle for the lattice inclination is formed due to an alignment of edge dislocation, the distance between the dislocations, $h$, may be $h=1.6$ nm from the relation as $h=b/\theta$, where $b$ is the Burgers vector and $\theta$ is the tilt angle. Because Mg atom is bigger and Cu atom is smaller than Al atom in diameter, they will relax the local strain around edge dislocations as illustrated in Fig. 12 where the dashed lines and solid lines are before and after the segregation of solute atoms, respectively. Note that the solute atoms are arbitrarily located in Fig. 12 so that the lattice becomes aligned with less strain and thus the dislocations becomes more stable. The dislocation movement through such a tilt boundary requires more energy as it creates lattice distortion, and it is reasonable to consider that the dislocation movement was impeded by the boundaries even they are low angles. The segregation of solute atoms on GBs prevents the dislocation movement, and it is then necessary to take into account the contribution to such strengthening even if they are segregated on LAGBs.

Several reports demonstrate the contribution of GB segregation to the strengthening. Molecular dynamic simulation by Vo *et al*. [50] suggested that the segregation of Nb on GBs in nanograined Cu reduced the grain boundary energy and gave rise to higher strength equivalent to the one close to the ideal strength. Zhao *et al*. [51] also showed using the first-principles calculation that Cu segregation on Al grain boundaries increases the grain boundary strength, suggesting that dislocations are more effectively blocked by the GBs. Wang *et al*. [52] conducted a finite element analysis compared with experimental results for a SUS316 stainless steel, and found that the segregation of solute atoms on subgrain boundaries (*i.e.*, LAGBs) with a network structure well blocks the dislocation movement. This then deduces that the subgrain size with segregation should be applicable to the Hall-Petch relation rather than the grain size with high-angle misorientations.

### 4.2.3 Grain boundary fraction with solute segregation

In order to make a rigorous estimation for the contribution of GBs with high-angle and low-angle misorientations to the total strengthening, the grain size should be re-evaluated in terms of grain boundaries with solute segregation. The volume fraction of GBs ($f$) is related to the grain size ($d$) through the following equation:

$$f = A d^{-\beta} \quad , \tag{6}$$

where $A$ and $\beta$ are constants depending on the width of GB. Here, this form of the relation has been derived as described in [Appendix](). We define $f_{all}$ and $f_{gb}$ as volume fractions of all GBs and of the GBs that contribute to the strengthening:

$$f_{all} = f_{HAGB} + f_{LAGB} \quad , \tag{7}$$

$$f_{gb} = f_{HAGB} + \kappa_{seg} \times f_{LAGB} \quad , \tag{8}$$

where $f_{HAGB}$ and $f_{LAGB}$ are the volume fractions of high-angle and low-angle GBs, and $\kappa_{seg}$ is the fraction of segregated GBs. This expression then involves a condition that the segregation occurs on all HAGBs and contributes to the strengthening together with the LAGBs with the fraction of $\kappa_{seg}$. Thus, the effective grain size for strengthening ($d_{eff}$) may be approximated as

$$d_{eff} = A^{1/\beta} f_{gb}^{-1/\beta} \quad . \tag{9}$$

Putting eq.(7) and eq.(8) in eq.(9), $d_{eff}$ is expressed by

$$d_{eff} = \left(\frac{1-\kappa_{seg}}{d_{HAGB}^{\beta}} + \frac{\kappa_{seg}}{d_{all}^{\beta}}\right)^{-1/\beta} \quad , \tag{10}$$

under the condition that the constants $A$ and $\beta$ are independent of the grain boundary types (*i.e.*, high-angle or low-angle misorientation). The value of $d_{eff}$ is then determined through eq.(10) from the measurements of $\kappa_{seg}$, $d_{HAGB}$ and $d_{all}$. In this study, the value of $\kappa_{seg}$ was used as 0.1 and 1 (from STEM and APT results) for the as-HPT-processed and subsequently peak-aged samples, respectively. Likewise, the values of 130 nm and 160 nm for $d_{HAGB}$, and 20 nm (from APT results) for $d_{all}$ were input, leading to the grain boundary strengthening of $\Delta\sigma_{gb(eff)}$ = 274$\pm$40 MPa and 565 MPa for the as-HPT-processed and the subsequently peak-aged states, respectively. The strengthening is achieved almost twice by aging after HPT processing. Zhao *et al.* showed that the average grain size obtained by a harmonic form well represents the Hall-Petch relation when grains are present in a bimodal distribution [53]. Eq.(10) is also a harmonic form except that it involves the constant $\beta$.

### 4.2.4 Total strengthening

We now evaluate the total strength by adding the contributions of cluster formation and segregation onto GBs including LAGBs. The total strengths are of 845$\pm$51 MPa and 904~1009$\pm$8 MPa for the as-HPT-processed and the subsequently peak-aged samples as shown in Table 2 ($\Delta\sigma_y$ (with $\Delta\sigma_{gb(eff)}$)). It should be noted that the estimation of the strength after aging does not include extra hardening due to nucleation of dislocations within small grains as discussed by Huang *et al.* [54]. Nevertheless, these values are well in agreement with those measured by tensile tests as shown in Fig. 3(b). In particular, the increase due to grain boundary strengthening through the Hall-Petch relation is prominent and this demonstrates that the segregation of Cu and Mg atoms on LAGBs plays a significant role in strengthening the alloy to as high as 1 GPa. Then, the conclusion can be reached that, thanks to segregations of solute atoms, the strength of LAGBs becomes almost equivalent to HAGBs, suggesting that new method to achieve high strength in UFG alloys is successfully developed.

Ultra-high strengthening to 1 GPa in aluminum alloy was thus far attained by Liddicoat *et al*. in an A7075 alloy using HPT processing [15]. Valiev *et al*. reported ultra-high strengthening to ~950 MPa in an A7475 alloy and an Al-5.7wt.%Mg alloy [16]. Otherwise, there is no report in other aluminum alloy systems by the use of SPD technique. The present study has demonstrated that the highest record is attained in the 2xxx series of aluminum alloys by HPT processing.

**5. Conclusions**

1. A solution-treated A2024 (Al-Cu-Mg) alloy reduced the grain size to 130 nm by HPT processing for 10 revolutions at room temperature. The tensile strength of ~910 MPa was achieved by HPT processing.

2. The subsequent aging after HPT processing further increased the tensile strength as high as ~1 GPa, despite the grain growth to 160 nm and the decrease in dislocation density.

3. Transmission electron microscopy (TEM) and atom probe tomography (APT) revealed that precipitate particles were less present in the matrix. Instead, the analyses by TEM and APT showed that solute atoms such as Cu and Mg were segregated on grain boundaries including low-angle grain boundaries.

4. The mechanism for the high strengthening achieved in this study was clarified in terms of the factors such as work hardening, solid-solution hardening, cluster hardening, precipitation hardening and grain boundary hardening. The analyses by TEM and APT including X-ray diffraction suggested that the segregation of Cu and Mg solute atoms not only onto high-angle grain boundaries but also onto low-angle grain boundaries plays an important role for the enhancement of the tensile strength.

**Acknowledgements**

One of the authors (TM) would like to acknowledge the Research Fellowship for Young Scientists from the Japan Society for Promotion of Science (JSPS) (No. JP16J07050 and No. JP19J01767). This work was supported in part by the Light Metals Educational Foundation of Japan, and in part by Grant-in-Aids for Scientific Research (S) and (A) from JSPS (No. JP26220909 and No. JP19H00830). HPT was carried out in the International Research Center on Giant Straining for Advanced Materials (IRC-GSAM) at Kyushu University. The authors would like to thank the Agence Nationale de la Recherche (ANR) for financial support (PRASA project - ANR-15-CE08-0029).

**Appendix**

The volume fraction of grain boundaries is calculated through the following procedures. First, the grain shape is approximated to be the Kelvin's tetrakaidecahedron (inset of Fig. A) with which three-dimensional arrangement is feasible without any clearance between the grains. Second, the segregated grain boundary width is set 2 nm which is estimated from the composition profiles obtained by APT analyses. Third, the volume fraction of grain boundaries is calculated as a function of the grain size which is taken as an average diameter of the Kelvin's tetrakaidecahedron. The relation is then plotted in Fig. A. Fourth, this relation is expressed through the form as

$$f = Ad^{-\beta} .$$

Fifth, the values of $A$ and $\beta$ are adjusted and the best fit is achieved with $A$=4.3 and $\beta$=0.97. These constant values are then used in this study

**Data availability**

The raw/processed data required to reproduce these findings cannot be shared at this time as the data also forms part of an ongoing study.

**References**

[1] A. Wilm: Physikalisch-metallurgische Untersuchungen über magnesiumhaltige Aluminiumlegierungen, Metallurgie, 8 (1911) 225-227.

[2] W. S. Miller, L. Zhuang, J. Bottema, A. J. Wittebrood, P. De Smet, A. Haszler and A. Vieregge: Recent development in aluminium alloys for the automotive industry, Mater. Sci. Eng. A 280 (2000) 37-49, https://doi.org/10.1016/S1003-6326(14)63305-7

[3] T. Dursun and C. Soutis: Recent developments in advanced aircraft aluminium alloys, Mater. Design 56 (2014) 862-871, https://doi.org/10.1016/j.matdes.2013.12.002

[4] E.O. Hall: The deformation and ageing of mild steel: III Discussion of results, Proc. Phys. Soc. B, 64 (1951) 747-753.

[5] N.J. Petch: The Cleavage Strength of Polycrystals, J. Iron Steel Inst., 174 (1953) 25-28.

[6] R.Z. Valiev, Y. Estrin, Z. Horita, T.G. Langdon, M.J. Zehetbauer and Y.T. Zhu: Producing bulk ultrafine-grained materials by severe plastic deformation, JOM 58 (2006) 33-39, https://doi.org/10.1007/s11837-006-0213-7.

[7] V. M. Segal, V. I. Reznikov, A. E. Drobyshevskiy and V. I. Kopylov: Plastic Working of Metals by Simple Shear, Russian Metall. 1 (1981) 99-105.

[8] P. W. Bridgman: Effects of High Shearing Stress Combined with High Hydrostatic Pressure, Phys. Rev. 48 (1935) 825-847, https://doi.org/10.1103/PhysRev.48.825.

[9] Y. Saito, H. Utsunomiya, N. Tsuji and T. Sakai: Novel ultra-high straining

process for bulk materials-development of the accumulative roll-bonding (ARB) process, Acta Mater. 47 (1999) 579–583, https://doi.org/10.1016/S1359-6454(98)00365-6.

[10] O.R. Valiakhmetov, R.M. Galeyev and G.A. Salishchev: Mechanical properties of the titanium alloy VT8 with submicrocrystalline structure, Fiz. Metal. Metalloved 10 (1990) 204-206.

[11] T. Fujioka and Z. Horita: Development of High-Pressure Sliding Process for Microstructural Refinement of Rectangular Metallic Sheets, Mater. Trans. 50 (2009) 930-933, https://doi.org/10.2320/matertrans.MRP2008445.

[12] S. D. Terhune, D. L. Swisher, K. Oishi, Z. Horita, T. G. Langdon and T. R. Mcnelley: An investigation of microstructure and grain-boundary evolution during ECA pressing of pure aluminum, Metall. Mater. Trans. A, 33 (2002) 2173-2184.

[13] Y. Ito and Z. Horita: Microstructural evolution in pure aluminum processed by high-pressure torsion, Mater. Sci. Eng. A, 503 (2009) 32-36.

[14] N. Tsuji, Y. Ito, Y. Saito and Y. Minamino: Strength and ductility of ultrafine grained aluminum and iron produced by ARB and annealing, Scr. Mater., 47 (2002) 893-899.

[15] P. V. Liddicoat, X. Z. Liao, Y. Zhao, Y. Zhu, M. Y. Murashkin, E. J. Lavernia, R. Z. Valiev and S. P. Ringer: Nanostructural hierarchy increases the strength of aluminium alloys, Nat. Commun. 1 (2010) 63(7p), https://www.nature.com/articles/ncomms1062.

[16] R. Z. Valiev, N. A. Enikeev, M. Yu. Murashkin, V. U. Kazykhanov and X. Sauvage: On the origin of the extremely high strength of ultrafine-grained Al alloys produced by severe plastic deformation, Scr. Mater. 63 (2010) 949-952.

[17] I. F. Mohamed, Y. Yonenaga, S. Lee, K. Edalati and Z. Horita: Age hardening and thermal stability of Al–Cu alloy processed by high-pressure torsion, Mater.

Sci. Eng. A 627 (2015) 111-118.

[18] I. F. Mohamed, T. Masuda, S. Lee, K. Edalati, Z. Horita, S. Hirosawa, D. Terada and M. Z. Omar: Strengthening of A2024 alloy by high-pressure torsion and subsequent aging, Mater. Sci. Eng. A 704 (2017) 112-118.

[19] Y. Tang, W. Goto, S. Hirosawa, Z. Horita, S. Lee, K. Matsuda and D. Terada: Concurrent strengthening of ultrafine-grained age-hardenable Al-Mg alloy by means of high-pressure torsion and spinodal decomposition, Acta Mater. 131 (2017) 57-64., https://doi.org/10.1016/j.actamat.2017.04.002

[20] W. J. Kim, C. S. Chung, D. S. Ma, S. I. Hong and H. K. Kim: Optimization of strength and ductility of 2024 Al by equal channel angular pressing (ECAP) and post-ECAP aging, Scr. Mater. 49 (2003) 333-338, https://doi.org/10.1016/S1359-6462(03)00260-4

[21] S. Lee, Z. Horita, S. Hirosawa and K. Matsuda: Age-hardening of an Al–Li–Cu–Mg alloy (2091) processed by high-pressure torsion, Mater. Sci. Eng. A 546 (2012) 82-89, https://doi.org/10.1016/j.msea.2012.03.029

[22] T. Masuda, Y. Takizawa, M. Yumoto, Y. Otagiri and Z. Horita: Extra Strengthening and Superplasticity of Ultra¬ne-Grained A2024 Alloy Produced by High-Pressure Sliding, Mater. Trans. 58 (2017) 1647-1655., https://doi.org/10.2320/matertrans.M2017242

[23] A. Deschamps, F. De Geuser, Z. Horita, S. Lee and G. Renou: Precipitation kinetics in a severely plastically deformed 7075 aluminium alloy, Acta Mater. 66 (2014) 105-117., https://doi.org/10.1016/j.actamat.2013.11.071

[24] Y. H. Zhao, X. Z. Liao, Z. Jin, R. Z. Valiev and Y. T. Zhu: Microstructures and mechanical properties of ultrafine grained 7075 Al alloy processed by ECAP and their evolutions during annealing, Acta Mater. 52 (2004) 4589-4599., https://doi.org/10.1016/j.actamat.2004.06.017

[25] K. S. Ghosh, N. Gao and M. J. Starink: Characterisation of high pressure

torsion processed 7150 Al–Zn–Mg–Cu alloy, Mater. Sci. Eng. A 552 (2012) 164-171., https://doi.org/10.1016/j.msea.2012.05.026

[26] S. P. Ringer, K. Hono, I. J. Polmear and T. Sakurai: Precipitation processes during the early stages of ageing in A1-Cu-Mg alloys, Appl. Surf. Sci., 94/95 (1996) 253-260., https://doi.org/10.1016/0169-4332(95)00383-5.

[27] Y. Nasedkina, X. Sauvage, E. V. Bobruk, M. Y. Murashkin, R. Z. Valiev and N. A. Enikeev: Mechanisms of precipitation induced by large strains in the Al-Cu system, J. Alloys Comp., 710 (2017) 736-747. https://doi.org/10.1016/j.jallcom.2017.03.312

[28] W. Xu, X.C. Liu, X.Y. Li and K. Lu: Deformation induced grain boundary segregation in nanolaminated Al–Cu alloy, Acta Materialia 182 (2020) 207–214, https://doi.org/10.1016/j.actamat.2019.10.036

[29] X. Sauvage, A. Duchaussoy and G. Zaher: Strain induced segregations in severely deformed materials, Mat. Trans. 60 (7) (2019) 1151-1158, https://doi.org/10.2320/matertrans.MF201919

[30] G. Sha, R. K. W. Marceau, X. Gao, B. C. Muddle and S. P. Ringer: Nanostructure of aluminium alloy 2024: Segregation, clustering and precipitation processes, Acta Mater. 59 (2011) 1659-1670, https://doi.org/10.1016/j.actamat.2010.11.033.

[31] A. Alhamidi and Z. Horita: Grain refinement and high strain rate superplasticity in alumunium 2024 alloy processed by high-pressure torsion, Mater. Sci. Eng. A 622 (2015) 139-145, https://doi.org/10.1016/j.msea.2014.11.009.

[32] Y. Chen, N. Gao, G. Sha, S. P. Ringer and M. J. Starink: Strengthening of an Al–Cu–Mg alloy processed by high-pressure torsion due to clusters, defects and defect–cluster complexes, Mater. Sci. Eng. A 627 (2015) 10-20, https://doi.org/10.1016/j.msea.2014.12.107.

[33] Y. Chen, N. Gao, G. Sha, S. P. Ringer and M. J. Starink: Microstructural evolution, strengthening and thermal stability of an ultrafine-grained Al–Cu–Mg alloy, Acta Mater. 109 (2016) 202-212, https://doi.org/10.1016/j.actamat.2016.02.050.

[34] S. V. Dobatkin, E. N. Bastarache, G. Sakai, T. Fujita, Z. Horita and T. G. Langdon: Grain refinement and superplastic flow in an aluminum alloy processed by high-pressure torsion, Mater. Sci. Eng. A 408 (2005) 141-146, https://doi.org/10.1016/j.msea.2005.07.023.

[35] G.K. Williamson and W.H. Hall: X-ray line broadening from filed aluminium and wolfram, Acta Metall. 1 (1953) 22-31, https://doi.org/10.1016/0001-6160(53)90006-6.

[36] G. K. Williamson and R. E. Smallman: III. Dislocation densities in some annealed and cold-worked metals from measurements on the X-ray debye-scherrer spectrum, Philos. Mag. 8 (1956) 24-46, https://doi.org/10.1080/14786435608238074.

[37] B. E. Warren and J. Biscoe: The Structure of Silica Glass by X-Ray Diffraction Studies, J. Am. Ceram. Soc., 21 (1938) 49-54. https://doi.org/10.1111/j.1151-2916.1938.tb15742.x.

[38] Y. Nasedkina, X. Sauvage, E. V. Bobruk, M. Y. Murashkin, R. Z. Valiev and N. A. Enikeev: Mechanisms of precipitation induced by large strains in the Al-Cu system, J. Alloys Comp., 710 (2017) 736-747. https://doi.org/10.1016/j.jallcom.2017.03.312

[39] A. Biswas, D. J. Siegel, C. Wolverton and D. N. Seidman: Precipitates in Al–Cu alloys revisited: Atom-probe tomographic experiments and first-principles calculations of compositional evolution and interfacial segregation, Acta Mater., 59 (2011) 6187-6204. https://doi.org/10.1016/j.actamat.2011.06.036

[40] I. Sabirov, M. Y. Murashkin and R. Z. Valiev: Nanostructured aluminium

alloys produced by severe plastic deformation: New horizons in development, Mater. Sci. Eng. A, 560 (2013) 1-24., https://doi.org/10.1016/j.msea.2012.09.020.

[41] N. Hansen: The effect of grain size and strain on the tensile flow stress of aluminium at room temperature, Acta Metall., 25 (1977) 863-869, https://doi.org/10.1016/0001-6160(77)90171-7.

[42] R.L. Fleischer: Substitutional solution hardening, Acta Metall., 11 (1963) 203–209, https://doi.org/10.1016/0001-6160(63)90213-X.

[43] R. Labusch: Statistical theories of solid solution hardening, Acta Metall., 20 (1972) 917–927, https://doi.org/10.1016/0001-6160(72)90085-5.

[44] J. E. Bailey and P. B. Hirsch: The dislocation distribution, flow stress, and stored energy in cold-worked polycrystalline silver, Philos. Mag., 5 (1960) 485–497, https://doi.org/10.1080/14786436008238300.

[45] N. Hansen and X. Huang: Microstructure and flow stress of polycrystals and single crystals, Acta Mater., 46 (1998) 1827-1836, https://doi.org/10.1016/S1359-6454(97)00365-0.

[46] G. I. Taylor: Plastic Strain in Metals, J. Inst. Met., 62 (1938), 307-324.

[47] G. Y. Chin and W. L. Mammel: Computer Solutions of the Taylor Analysis for Axisymmetric Flow, Trans. of the Met. Soc. of AIME, 239 (1967) 1400-1405.

[48] B. Clausen: Characterisation of Polycrystal Deformation by Numerical Modelling and Neutron Diffraction Experiments (1997).

[49] T. Shanmugasundaram, M. Heilmaier, B. S. Murty and V. S. Sarma: On the Hall–Petch relationship in a nanostructured Al–Cu alloy, Mater. Sci. Eng. A, 527 (2010) 7821-7825, https://doi.org/10.1016/j.msea.2010.08.070.

[50] N. Q. Vo, J. Schäfer, R. S. Averback, K. Albe, Y. Ashkenazy and P. Bellon: Reaching theoretical strengths in nanocrystalline Cu by grain boundary doping, Scr. Mater., 65 (2011) 660-663.,

https://doi.org/10.1016/j.scriptamat.2011.06.048.

[51] D. Zhao, O. M. Løvvik, K. Marthinsen and Y. Li: Segregation of Mg, Cu and their effects on the strength of Al Σ5 (210)[001] symmetrical tilt grain boundary, Acta Mater., 145 (2018) 235-246., https://doi.org/10.1016/j.actamat.2017.12.023.

[52] Y. M. Wang, T. Voisin, J. T. McKeown, J. Ye, N. P. Calta, Z. Li, Z. Zeng, Y. Zhang, W. Chen, T. T. Roehling, R. T. Ott, M. K. Santala, P. J. Depond, M. J. Mattews, A. V. Hamza and T. Zhu: Additively manufactured hierarchical stainless steels with high strength and ductility, Nat. Mater., 17 (2018) 63-71., https://doi.org/10.1038/nmat5021.

[53] M. Zhao, F. Yin, T. Hanamura, K. Nagai and A. Atrens: Relationship between yield strength and grain size for a bimodal structural ultrafine-grained ferrite/cementite steel, Scr. Mater., 57 (2007) 857-860, https://doi.org/10.1016/j.scriptamat.2007.06.062.

[54] X. Huang, N. Hansen and N. Tsuji: Hardening by Annealing and Softening by Deformation in Nanostructured Metals, Science, 312 (2006) 249-251, https://doi.org/10.1126/science.1124268

**Figure and table captions**

**Table 1** Compositions measured from APT data of Fig. 10. About 0.5at.% of solute (Cu plus Mg) has segregated along boundaries after processing by HPT ($N$=10) and aging for 35 min at 423 K.

**Table 2** Contribution of strengthening factors estimated using theoretical approach, and comparison with theoretical values for as-HPT-processed and subsequently peak-aged states.

**Fig. 1** Sample dimensions of HPT disk with positions marked for Vickers microhardness measurement, position extracted for tensile specimen with dimensions, and location of TEM specimen.

**Fig. 2** (a) Stress-strain curves and (b) variation of tensile strength and elongation to failure obtained after deformation at R.T. with initial strain rate of $3.0\times10^{-3}$ s$^{-1}$ after $N$= 0.75, 1, 3, 5 and 10 including ST and T6 sample.

**Fig. 3** (a) Hardness variations plotted against corresponding to equivalent strain and (b) stress-strain curves obtained after deformation at R.T. with initial strain rate of $3.0\times10^{-3}$ s$^{-1}$ for the sample after $N$=10 and subsequent peak-aging including ST and T6 sample.

**Fig. 4** TEM bright-field images (left), dark-field images (right) and selected area electron diffraction patterns (inset) for sample after (a) $N$=10 and (b) subsequent peak-aging.

**Fig. 5** XRD profiles for samples after $N$=10 and subsequent peak-aging (35 min) and over-aging (3 h) including ST sample.

**Fig. 6** STEM microstructures for sample after (a,b) $N$=10 and (c,d) subsequent peak-aging, and (e) intensity variation of HAADF signal and (f) EDS line profiles obtained along $x$ in (c,d).

**Fig. 7** Reconstructed three-dimensional atom maps of (a) Al, Cu and Mg, (b) Cu and (c) Mg obtained by APT analysis for as-HPT processed sample.

Concentration profiles of Cu and Mg across GB (sampling volume thickness: 0.1 nm) are also shown in (d).

**Fig. 8** Reconstructed three-dimensional atom maps of (a) Cu and (b) Mg obtained by APT analysis for subsequently peak-aged sample.

**Fig. 9** Three-dimensional reconstruction of other volume analyzed by APT for subsequently peak-aged sample; (a) distribution of Al and Cu, (b) distribution of Al and Mg, same volume viewed in another direction and filtered to exhibit only atoms where local compositions exceed (c) 4at.%Cu and (d) 15at.%Cu and 4at.%Mg.

**Fig. 10** Typical frequency distribution histograms of Cu and Mg computed far from segregated region in (a) as-HPT-processed (Fig. 7) and (b) subsequently peak-aged samples (Fig. 9). Frequency distribution histograms of measured data are compared to histograms of same data set where atom positions were randomized (sampling volume 2x2x2 nm$^3$). Concentration profiles computed across two GBs (sampling volume thickness: 0.1 nm) in Fig. 9 are also shown in (c).

**Fig. 11** Schematic illustration of microstructures in terms of solute atoms for (a) as-HPT-processed and (b) subsequently peak-aged states.

**Fig. 12** Schematic illustration of segregated subgrain boundary with 10° of misorientation.

**Fig. A** Relationship between average grain size and volume fraction of grain boundary.

Table 1   Compositions measured from APT data of Fig. 10. About 0.5at.% of solute (Cu plus Mg) has segregated along boundaries after processing by HPT (*N*=10) and aging for 35 min at 423 K.

| at.% | Whole data set | Grain interior |
|------|----------------|----------------|
| Cu   | 1.48 ± 0.01    | 1.23 ± 0.04    |
| Mg   | 1.42 ± 0.01    | 1.27 ± 0.04    |

Table 2   Contribution of strengthening factors estimated using theoretical approach, and comparison with theoretical values for as-HPT-processed and subsequently peak-aged states.

|  | As-HPT | HPT+peak-aging |
|---|---|---|
| $\sigma_0$ | ~5 MPa | ~5 MPa |
| $\Delta\sigma_{ss}$ | ~50 MPa | ~50 MPa |
| $\Delta\sigma_{dis}$ | 406 ± 11 MPa | 174 ± 8 MPa |
| $\Delta\sigma_{pre}$ | ~0 | ~0 |
| $\Delta\sigma_{cluster}$ | ~110 MPa | 110~215 MPa |
| $\Delta\sigma_{gb\ (HAGB)}$ | 222 ± 40 MPa | 200 ± 50 MPa |
| $\Delta\sigma_{gb\ (eff)}$ | 274 ± 40 MPa | 565 MPa |
| $\sigma_y(with\ \Delta\sigma_{gb\ (HAGB)})$ | 793 ± 51 MPa | 539~644 ± 58 MPa |
| $\sigma_y(with\ \Delta\sigma_{gb\ (eff)})$ | 845 ± 51 MPa | 904~1009 ± 8 MPa |
| $\sigma_y(measured)$ | 770 MPa | 955 MPa |

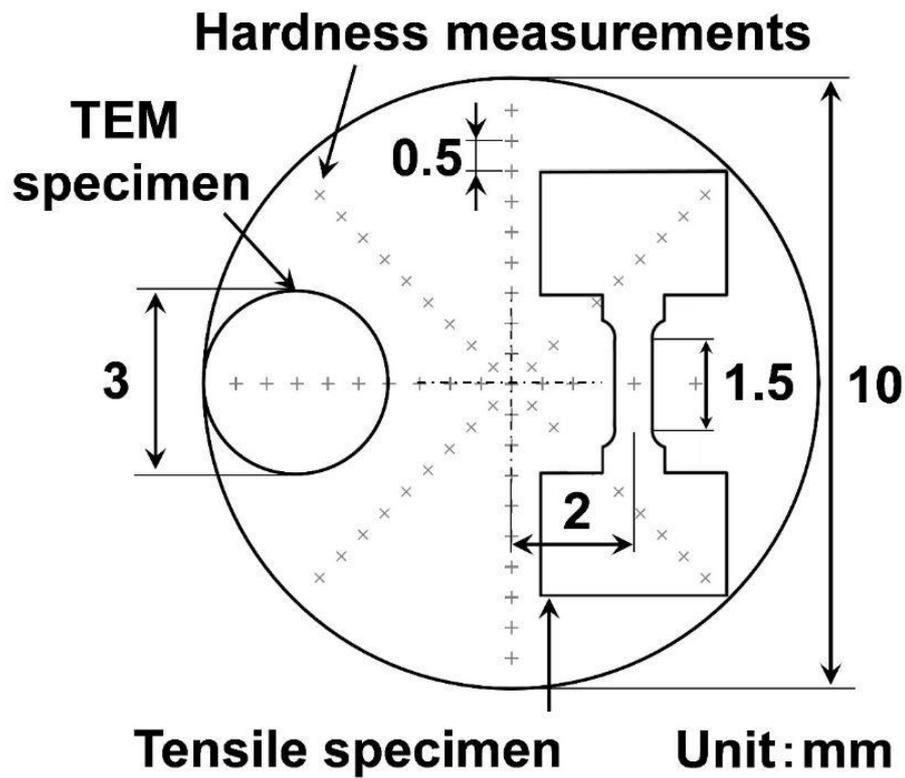

**Fig. 1** Sample dimensions of HPT disk with positions marked for Vickers microhardness measurement, position extracted for tensile specimen with dimensions, and location of TEM specimen.

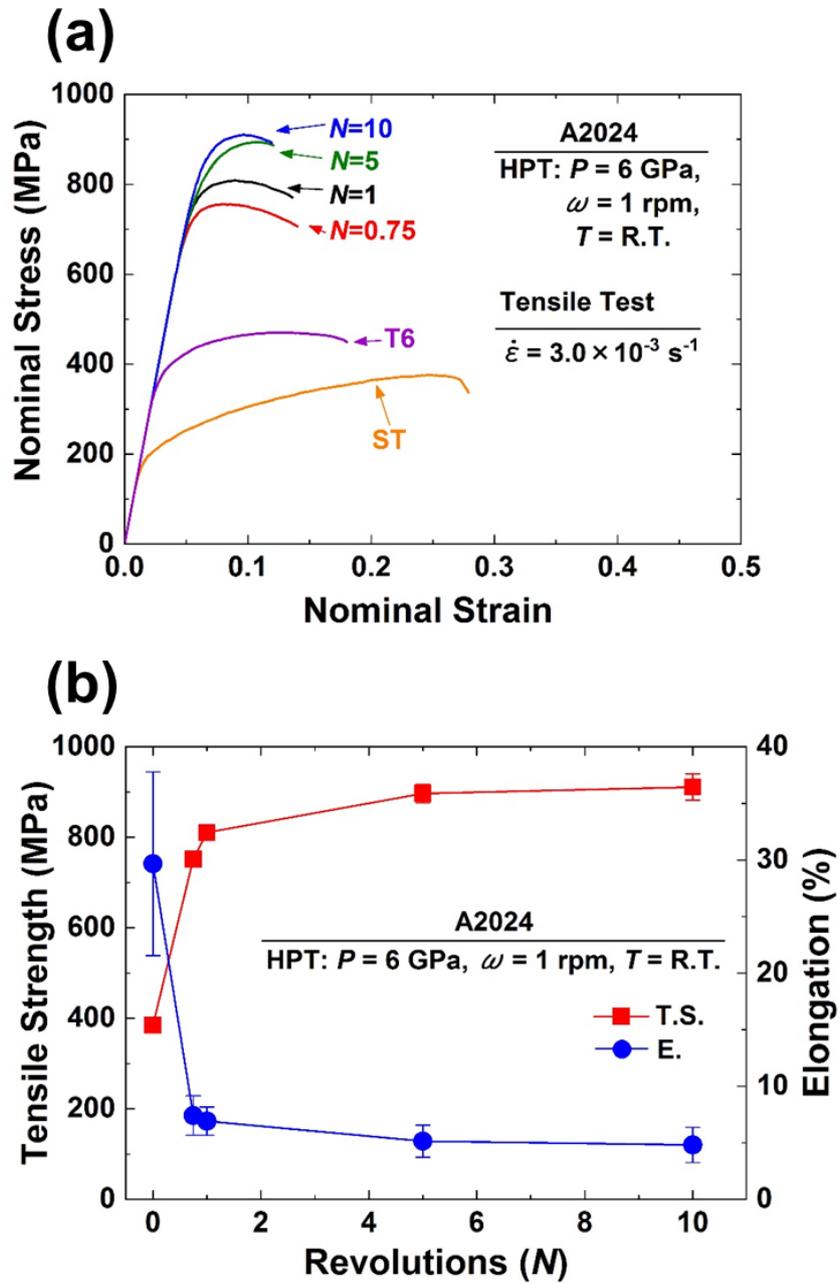

**Fig. 2** (a) Stress-strain curves and (b) variation of tensile strength and elongation to failure obtained after deformation at R.T. with initial strain rate of $3.0\times10^{-3}$ s$^{-1}$ after $N$= 0.75, 1, 3, 5 and 10 including ST and T6 sample.

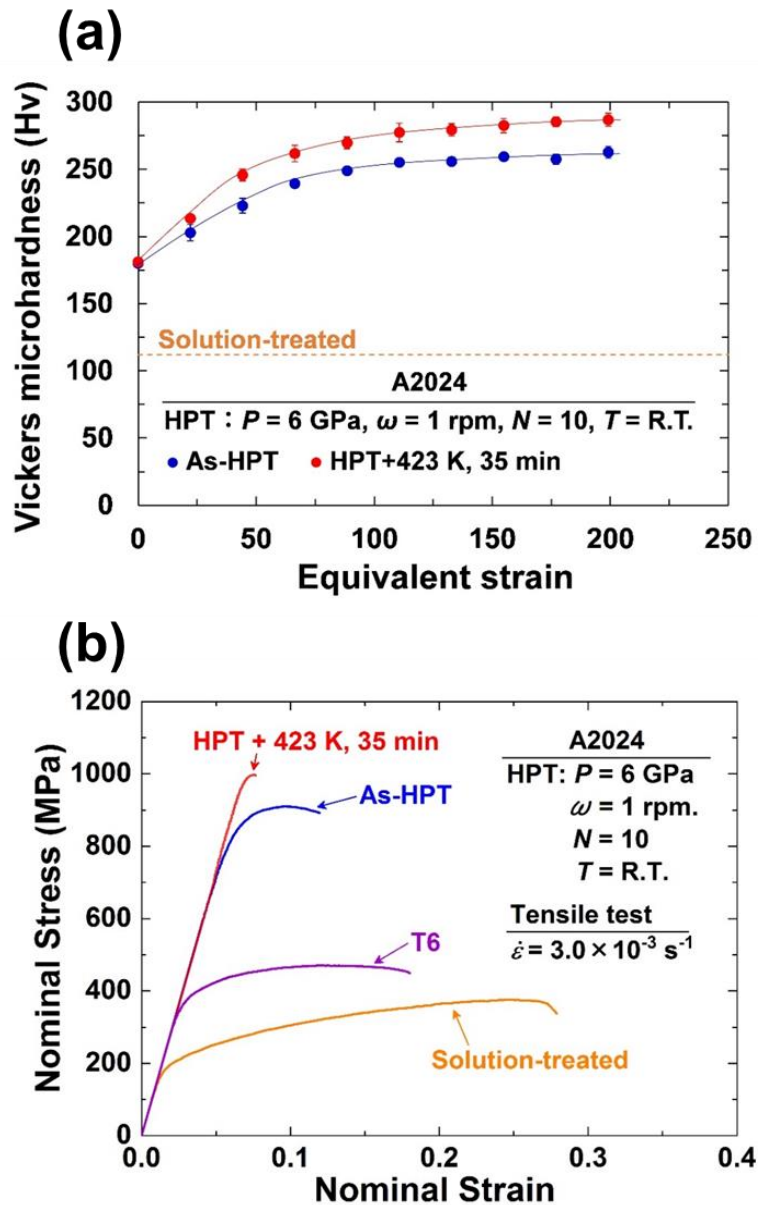

**Fig. 3** (a) Hardness variations plotted against corresponding to equivalent strain and (b) stress-strain curves obtained after deformation at R.T. with initial strain rate of $3.0 \times 10^{-3}$ s$^{-1}$ for the sample after $N$=10 and subsequent peak-aging including ST and T6 sample.

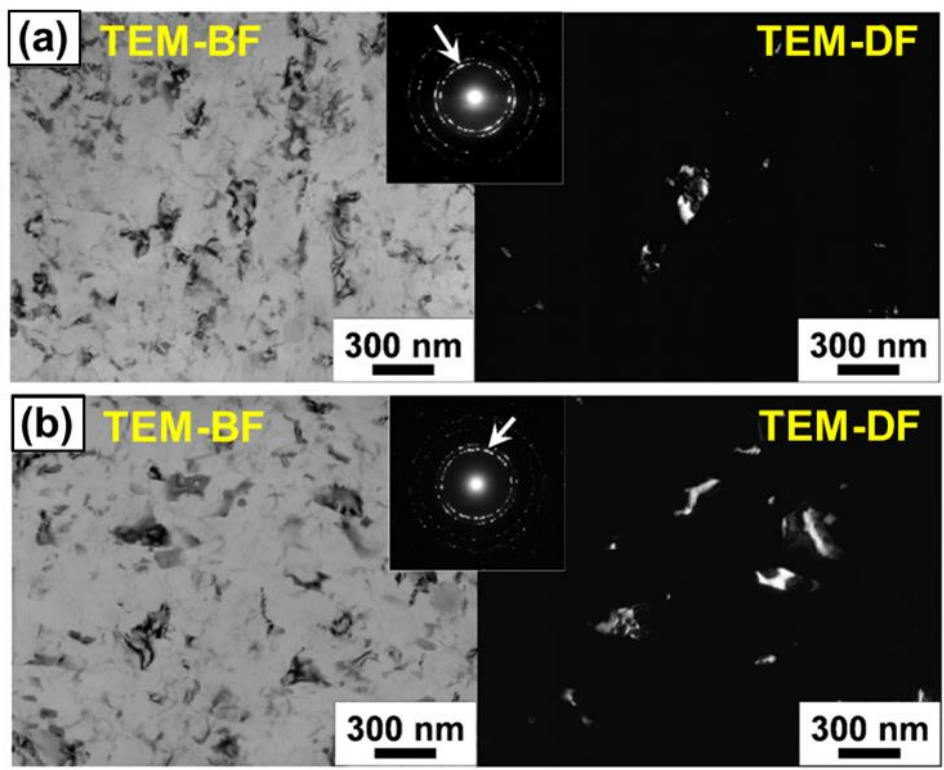

**Fig. 4** TEM bright-field images (left), dark-field images (right) and selected area electron diffraction patterns (inset) for sample after (a) $N$=10 and (b) subsequent peak-aging.

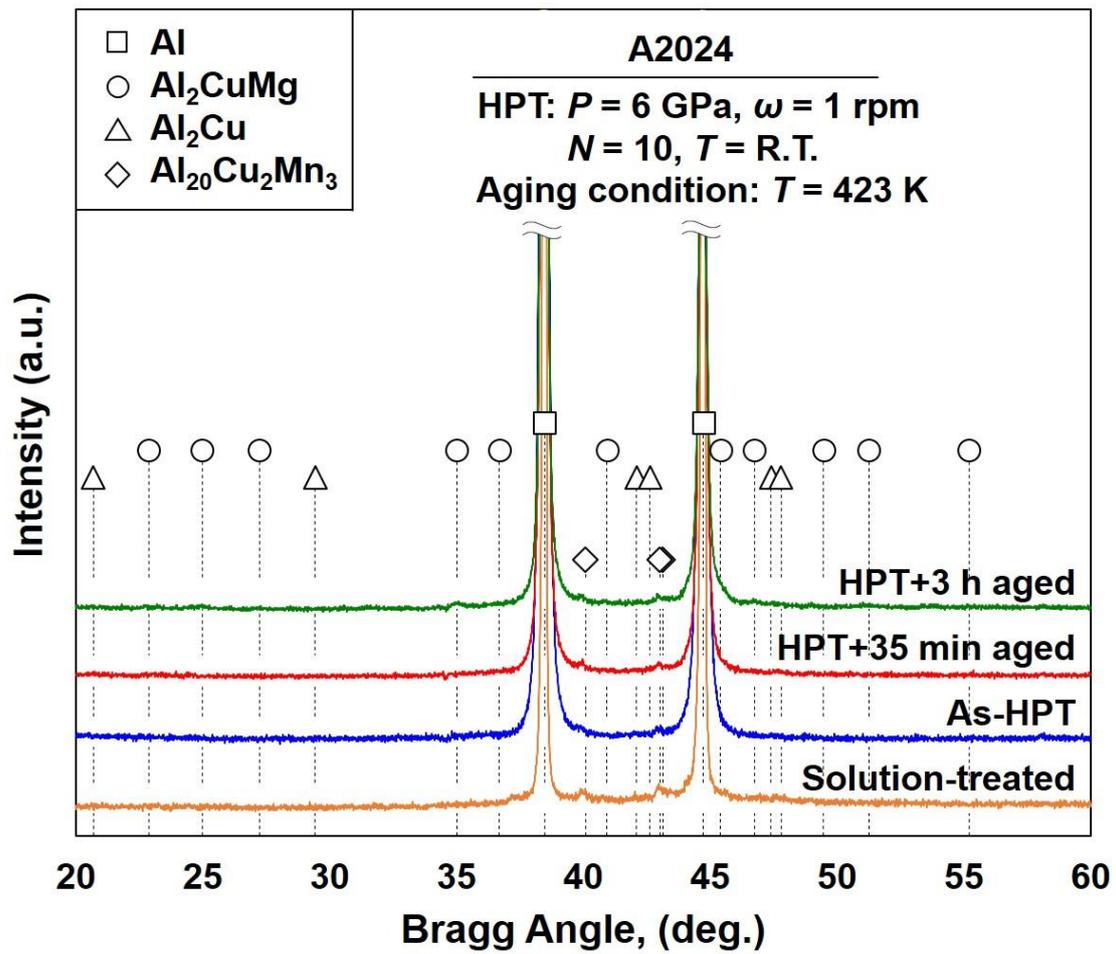

**Fig. 5** XRD profiles for samples after $N$=10 and subsequent peak-aging (35 min) and over-aging (3 h) including ST sample.

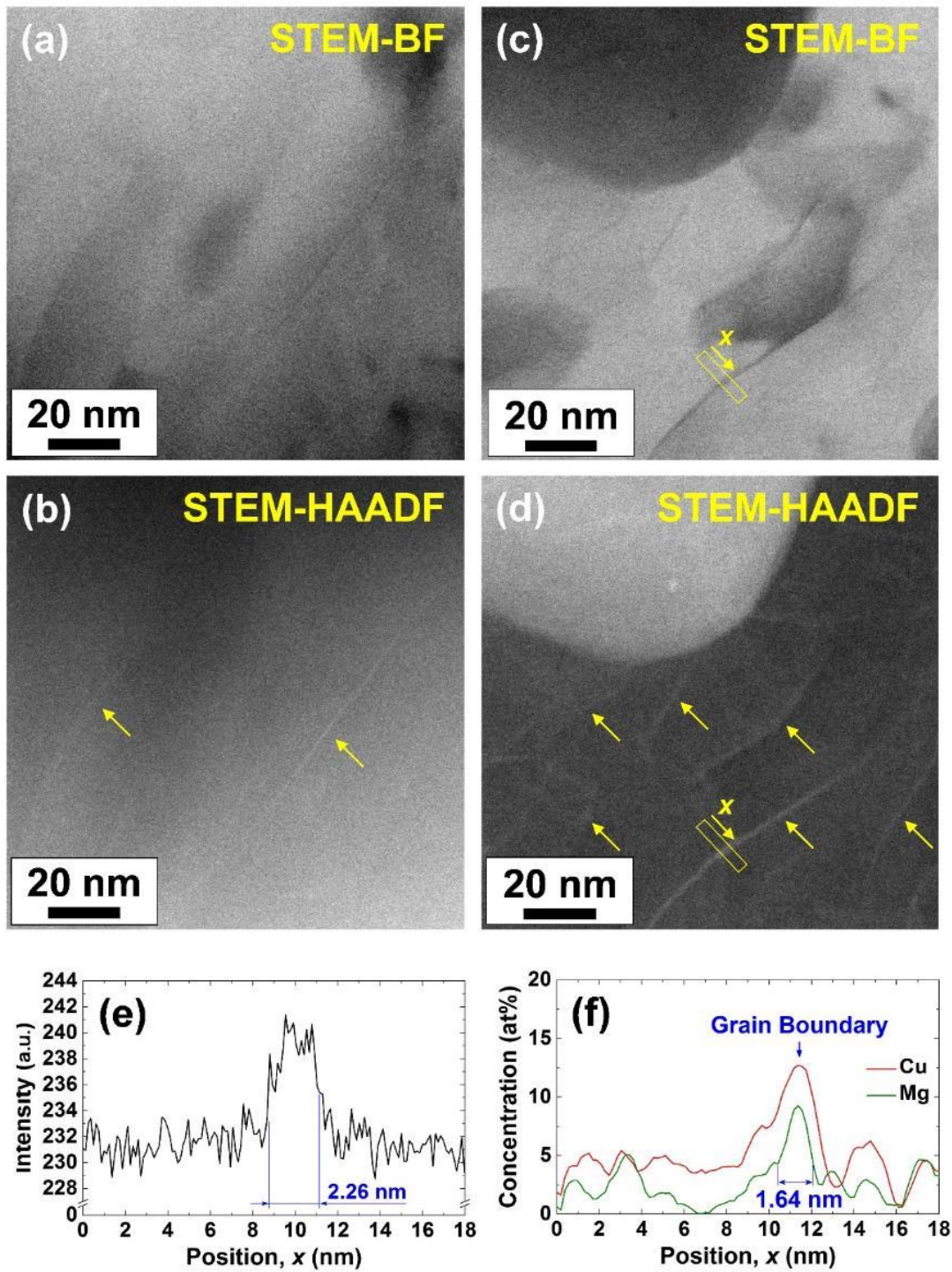

**Fig. 6** STEM microstructures for sample after (a,b) $N$=10 and (c,d) subsequent peak-aging, and (e) intensity variation of HAADF signal and (f) EDS line profiles obtained along $x$ in (c,d).

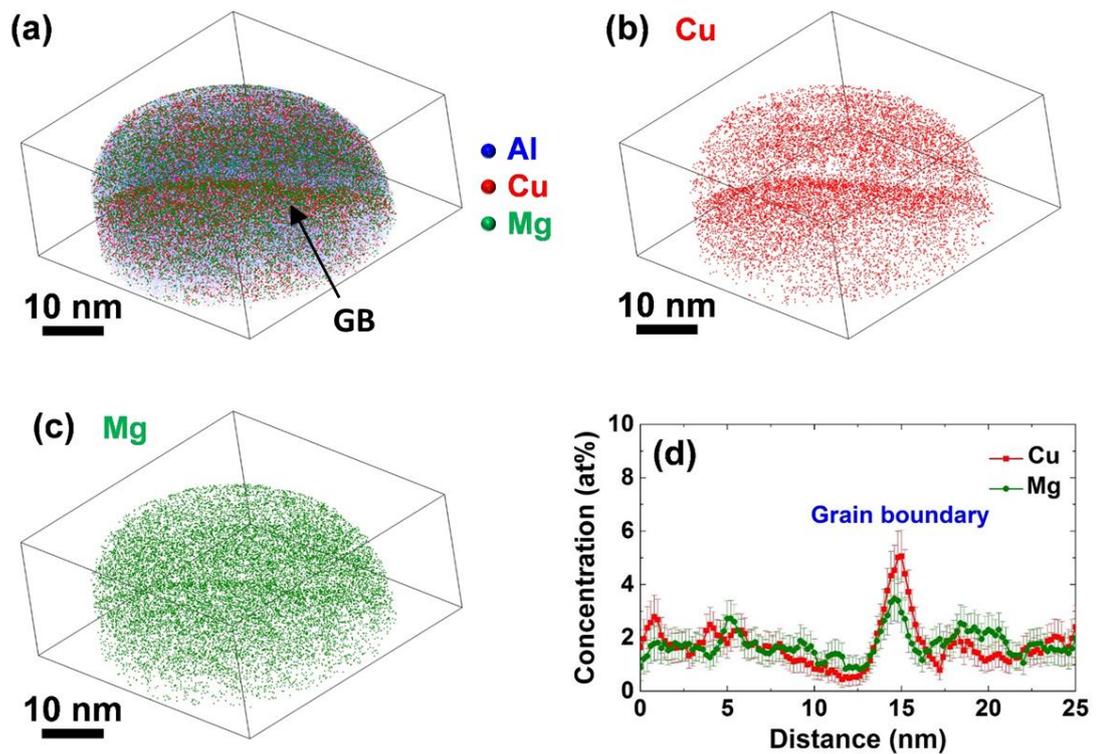

**Fig. 7** Reconstructed three-dimensional atom maps of (a) Al, Cu and Mg, (b) Cu and (c) Mg obtained by APT analysis for as-HPT-processed sample. Concentration profiles of Cu and Mg across GB (sampling volume thickness: 0.1 nm) are also shown in (d).

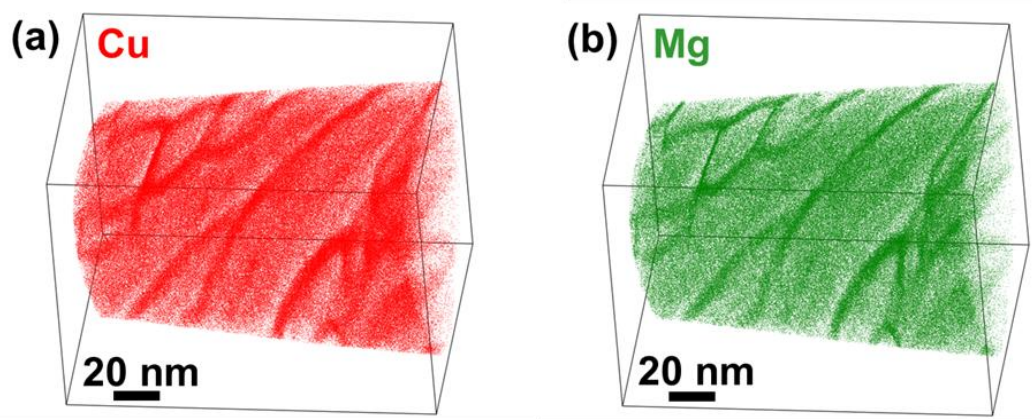

**Fig. 8** Reconstructed three-dimensional atom maps of (a) Cu and (b) Mg obtained by APT analysis for subsequently peak-aged sample.

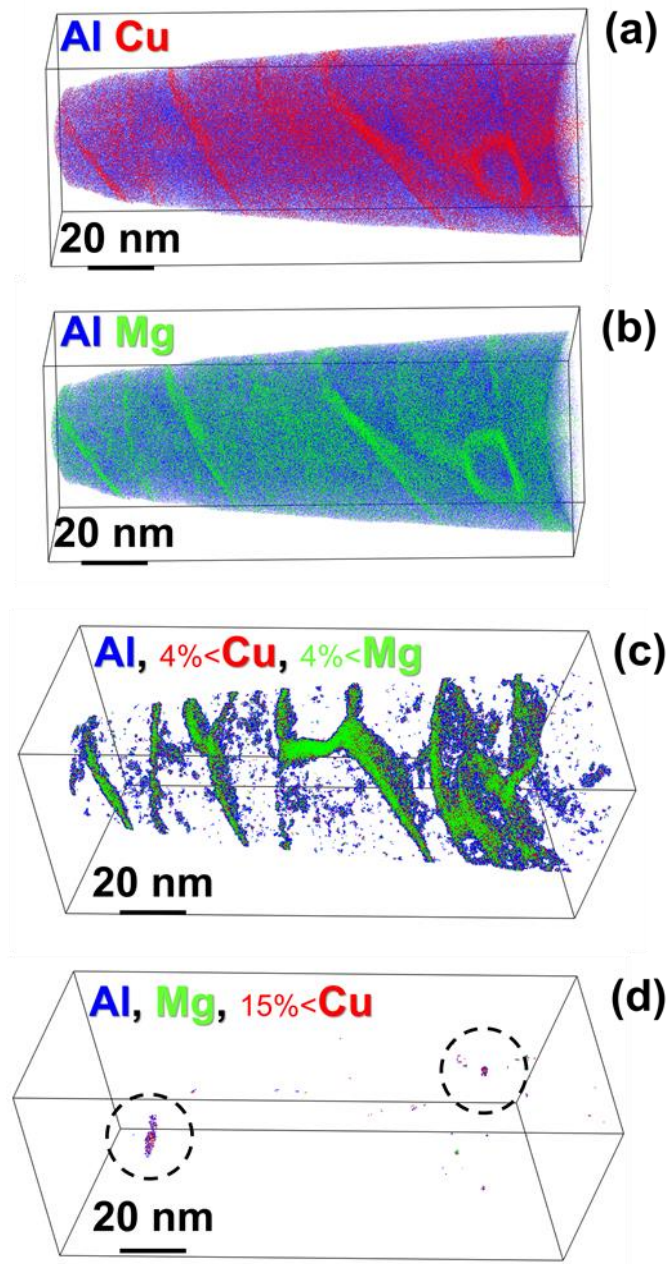

**Fig. 9** Three-dimensional reconstruction of other volume analyzed by APT for subsequently peak-aged sample; (a) distribution of Al and Cu, (b) distribution of Al and Mg, same volume viewed in another direction and filtered to exhibit only atoms where local compositions exceed (c) 4at.%Cu and (d) 15at.%Cu and 4at.%Mg.

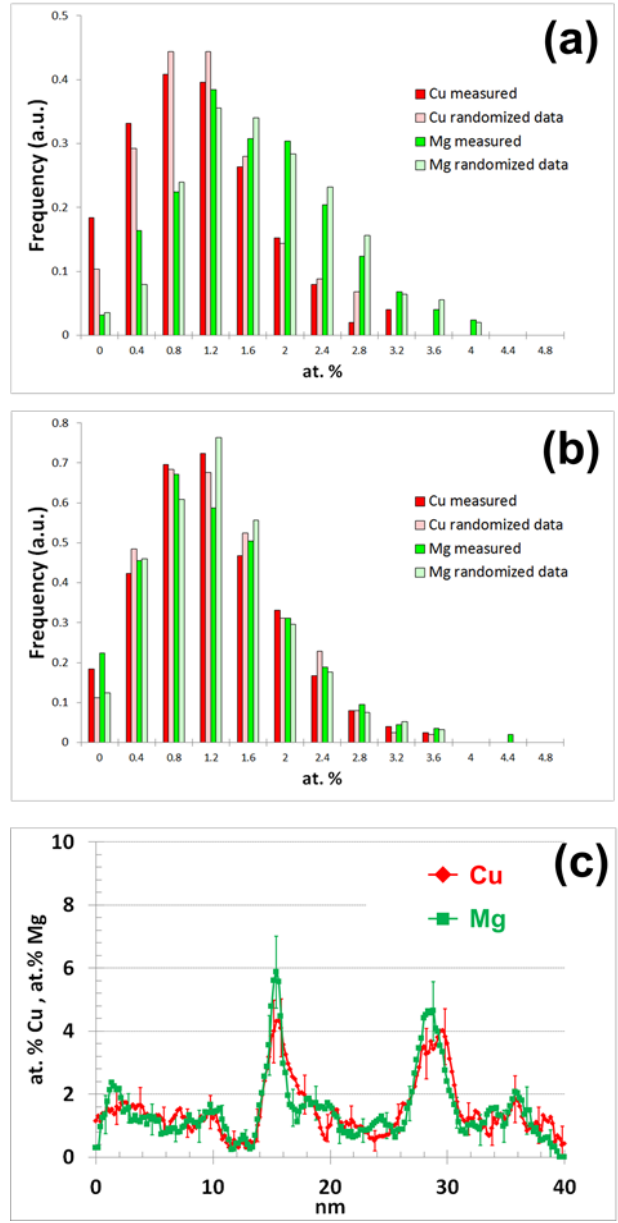

**Fig. 10** Typical frequency distribution histograms of Cu and Mg computed far from segregated region in (a) as-HPT-processed (Fig. 7) and (b) subsequently peak-aged samples (Fig. 9). Frequency distribution histograms of measured data are compared to histograms of same data set where atom positions were randomized (sampling volume 2x2x2 nm$^3$). Concentration profiles computed across two GBs (sampling volume thickness: 0.1 nm) in Fig. 9 are also shown in (c).

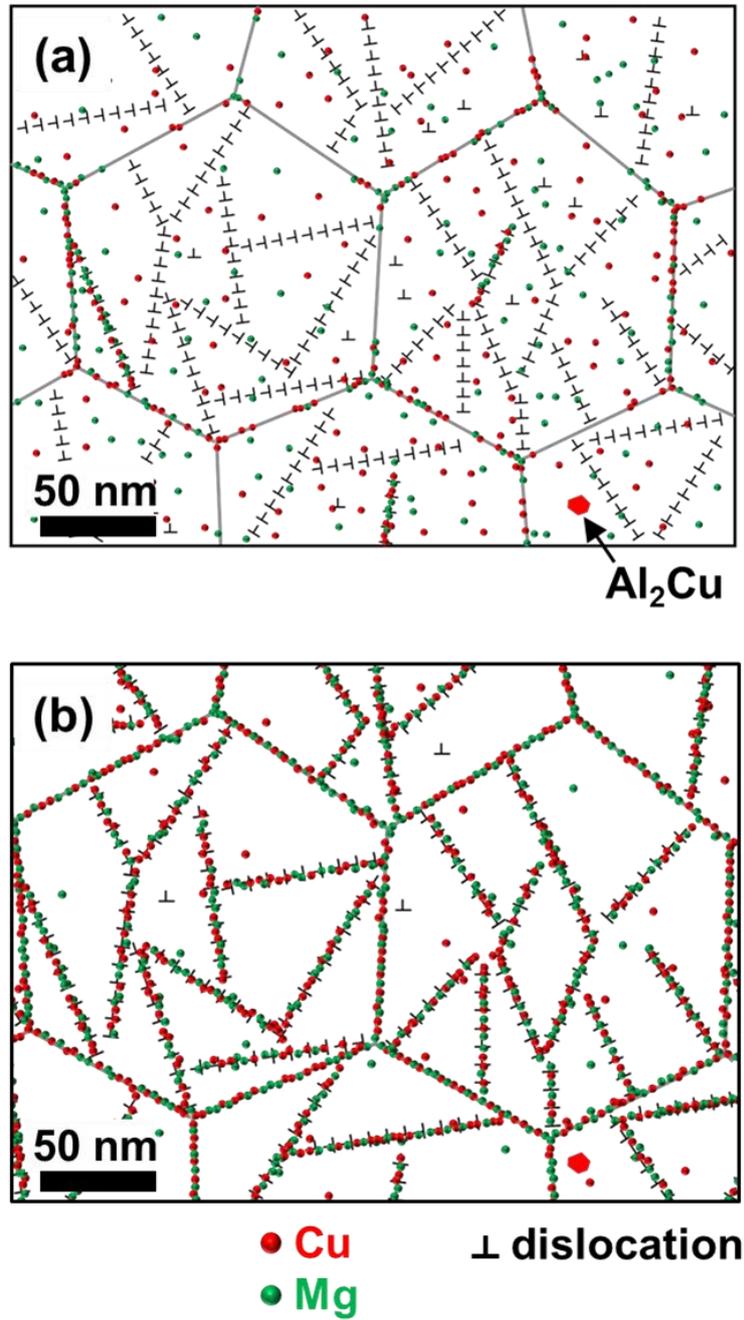

**Fig. 11** Schematic illustration of microstructures in terms of solute atoms for (a) as-HPT-processed and (b) subsequently peak-aged states.

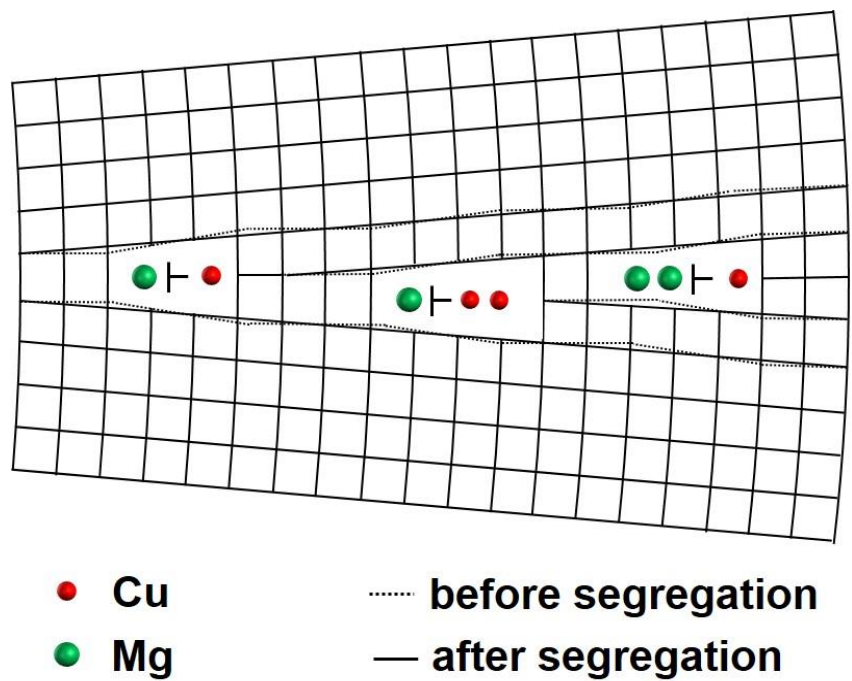

**Fig. 12** Schematic illustration of segregated subgrain boundary with 10° of misorientation.

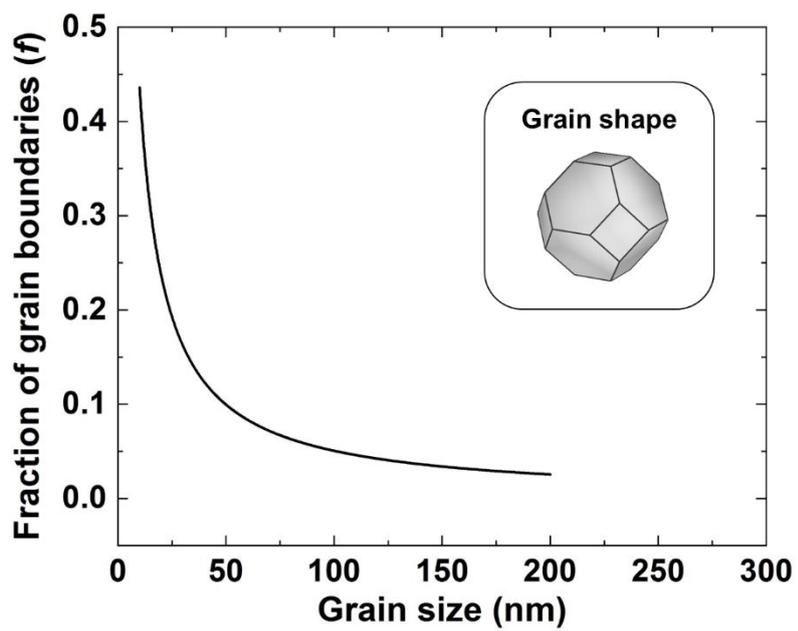

**Fig. A** Relationship between average grain size and volume fraction of grain boundaries.